\documentclass{appolb}
\usepackage{graphicx}
\usepackage{subfigure}
\def\cA{{\cal A}}

\def\cC{{\cal C}}

\def\qq{\overline{q}q}
\def\uu{\overline{u}u}
\def\dd{\overline{d}d}
\def\be{\begin{equation}}
\def\ee{\end{equation}}
\newcommand{\GS}[1]{#1\!\!\!\!\!\not~}
\newcommand{\tr}{\mbox{tr}}

\begin{document}
\title{Quark Condensates: Flavour Dependence%
\thanks{Contribution to the Proceedings of the EURICIDE Final Meeting, August 24-27th, 2006, Kazimierz, Poland}%
}
\author{R.~Williams
\address{Institute for Particle Physics Phenomenology, Durham University,\\ Durham  DH1 3LE, UK}
\and C.~S.~Fischer
\address{University of Technology Darmstadt, \\Schlossgartenstra\ss e 9, 64289 Darmstadt, Germany}
\and M.~R.~Pennington
\address{Institute for Particle Physics Phenomenology, Durham University,\\ Durham  DH1 3LE, UK}
}
\maketitle
\begin{abstract}
We determine the $\qq$ condensate for quark masses from zero up to that of the strange quark within a phenomenologically successful modelling of continuum QCD
by solving the quark Schwinger-Dyson equation. The existence of multiple solutions to this equation is the key to an accurate and reliable extraction of this condensate using the operator product expansion. 
We explain why alternative definitions fail to give the physical condensate.
\end{abstract}

\newpage
\section{Introduction}
The dynamics of low energy hadrons are governed by the non-trivial structure of the QCD vacuum and the resulting breaking of chiral symmetry. Long-range correlations between quarks and antiquarks form condensates, whose scale determines constituent quark masses and so the masses of all light hadrons. Though the scale has been recently determined for the $\uu$ and $\dd$ through experiments involving $\pi\pi$ interactions, confirming the $\sim\,-(235 \,{\rm MeV})^3$ anticipated from phenomenology~\cite{Pennington:2005be}, we are yet to determine a value of the condensate for the not-so-light quarks with any certainty.

The interest in the value of such a condensate arises in the context of QCD sum-rules where the Operator Product Expansion (OPE) is used to approximate the short distance behaviour of QCD. 
In studying currents $\,\overline{q_i} \gamma^{\mu} (\gamma_5) q_j$, with $q_i=s$ and $q_j=u,d$, the VEVs of $\uu$, $\dd$ and $\overline{s}s$ operators naturally arise~\cite{Jamin:2006tj,Jamin:2002ev,Jamin:2001fw}. 
The $\qq$ condensate for the $u$ and $d$ are expected to be close to that in the chiral limit, with even an error of 10\% unimportant in previous calculations. 
However, if one considers strange quarks we are left with the estimate of Shifman \emph{et al}~\cite{Shifman:1978bx,Shifman:1978by} that
the $\overline{s}s$ condensate be $(0.8\,\pm\,0.3)$ of the $\uu$ and $\dd$ values. 
It is the greater precision brought about by the studies of Refs.~\cite{Jamin:2006tj,Maltman:2001jx,Maltman:2002sb}, for instance, that motivate the need to learn about how the $\qq$ condensate depends on the current quark mass. Our goal here is to illustrate a method for determining this dependence.


\section{Schwinger--Dyson Equations}

\begin{figure}[b]
\vspace{1mm}
 \begin{center}  
       \includegraphics*[width=0.95\columnwidth]{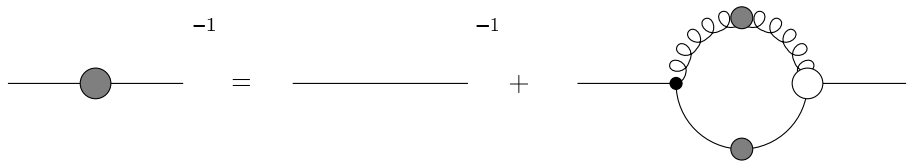}
	\caption{Schwinger-Dyson equation for the quark propagator}\label{quark:sde}
\end{center}  
  \end{figure} 
Our aim is to calculate the mass function of the quark propagator for a range of current masses. The starting point is the renormalized Schwinger-Dyson equation for the quark propagator as depicted in Fig.~\ref{quark:sde}:
\begin{eqnarray}\label{cond:eqn:quarksde1}	
\vspace{2mm}
	S_F^{-1}(p) &=&  Z_2\left[ S_F^{(0)}(p) \right]^{-1}-C_F \frac{\tilde{Z}_1\,Z_2}{\tilde{Z}_3}\frac{g^2}{\left( 2\pi \right)^4}\int d^4k \cdots\nonumber\\[2mm]
	&&\hspace{35pt}\times\gamma_\mu\, S_F(k)\,\Gamma_\nu(k,p)\,D_{\mu\nu}(p-k)\;.
\vspace{2mm}
\end{eqnarray}
In the Landau gauge we note that $\tilde{Z}_1 = 1$. The inverse propagator $S_F^{-1}(p)$ is specified by two scalar functions $\cA$ and $M$:
\begin{equation}\label{fermion:eqn:AB}
	S_F^{-1}(p)\,=\,\cA(p^2,\mu^2)\left(\GS{p} +\;M(p^2)\right)\; ,
\end{equation}
where the quark mass function $M(p^2)$ is renormalisation group invariant.

\subsection{Maris--Tandy Model}
To solve Eq.~(\ref{cond:eqn:quarksde1}) we employ some suitable ansatz for the coupling and interaction in Eq.~(\ref{cond:eqn:quarksde1}) which has sufficient integrated strength in the infrared to achieve dynamical mass generation. 
Following Maris {\it et al.}~\cite{Maris:1997hd,Maris:1998hc}, we employ an ansatz for $g^2D_{\mu\nu}(p-k)$ shown to be consistent with studies of bound state mesons. 
The renormalisation scheme is one of modified momentum subtraction at point $\mu$, taken to be $19$ GeV to compare with earlier studies~\cite{Chang:2006bm}.
 We later evolve $\mu$ to the more common 2 GeV scale in the $\overline{MS}$-scheme. We use:
\begin{eqnarray}
	\frac{g^2}{4\pi}\frac{Z_2}{\tilde{Z}_3}D_{\mu\nu}(q)\rightarrow \alpha\left(q^2\right) D^{(0)}_{\mu\nu}(q)
\end{eqnarray}

\noindent where the coupling is described by ~\cite{Maris:1997hd,Maris:1998hc}:
\begin{eqnarray}\label{eqn:modelparam}
	\alpha\left( q^2 \right) &=& \frac{\pi}{\omega^6}\,D\, q^4\, \exp(-q^2/\omega^2)+\frac{2\pi \gamma_m}{\log\left( \tau+\left(1+q^2/\Lambda_{QCD}^2 \right)^2\right)}\nonumber\\[0.mm]
	&&\hspace{3.7cm}\times \left[ 1-\exp\left(-q^2/\left[ 4m_t^2 \right]\right) \right]\;,
\end{eqnarray}
with  $m_t= 0.5$ GeV, $\tau=e^2-1$, $\gamma_m=12/(33-2N_f)$ and $\Lambda_{QCD}=0.234$ GeV. We choose $\omega$ and $D$ to be consistent with meson observables; a typical set is $\omega=0.4$ GeV, $D=0.933$ GeV$^2$.
Solutions are obtained by solving for $\cA$ and $M$ of Eq.~(\ref{fermion:eqn:AB}), which we may write symbolically as:
\begin{eqnarray}
	\cA(p^2,\mu)&=&Z_2(\mu,\Lambda) - \Sigma_D\left(p,\Lambda\right)\; ,\nonumber\\[-1mm]
&& \\[-1.mm]
	M(p^2)\cA(p^2,\mu)&=&Z_2(\mu,\Lambda) Z_m m_R(\mu) + \Sigma_S\left(p,\Lambda\right)\;.\nonumber
\end{eqnarray}
The $\Sigma_S$ and $\Sigma_D$ correspond to the 
scalar and spinor projections of the integral in Eq.~(\ref{cond:eqn:quarksde1}). For massive quarks we obtain the solution $M$ (later called $M^+$) by eliminating the renormalisation factors $Z_2$, $Z_m$ via:
\begin{eqnarray}
	Z_2(\mu,\Lambda)&=& 1+\Sigma_D\left( \mu,\Lambda \right)\; ,\nonumber\\[-2.mm]
  && \\[-2.mm]
	Z_m(\mu,\Lambda)&=& \frac{1}{Z_2(\mu,\Lambda)}-\frac{\Sigma_S\left(\mu,\Lambda \right)}{Z_2(\mu,\Lambda) m_R\left(\mu\right)}\;.\nonumber
\end{eqnarray}
The momentum dependence for different values of $m_R$ are shown in Fig.~\ref{masses}. Our purpose is to define the value of the $\qq$ condensate for each of these.

\section{Extracting the Condensate}
\begin{figure}[b!]
\vspace{1mm}
 \begin{center}  
       \includegraphics*[width=0.65\columnwidth]{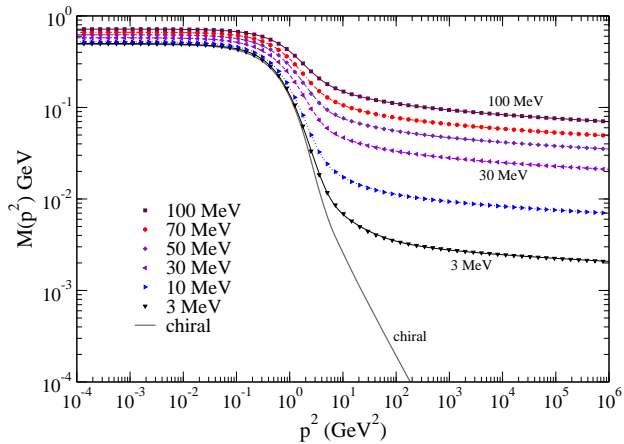}
	\caption{Euclidean mass functions for different current masses, specified at $\mu = 19$ GeV as labelled. The plot illustrates how on a log-log plot the behaviour dramatically changes between a current mass of 0 and 3~MeV.}\label{masses}
\end{center}  
  \end{figure} 
At very large momenta the tail of the mass function is described by the operator product expansion so that
\begin{eqnarray}\label{cond:eqn:opefit}
	M(p^2)_{asym} &=&  \overline{m}\left[ \log\left( p^2/\Lambda_1^{\,2} \right] \right)^{-\gamma_m}+\frac{2\pi^2\gamma_m}{3}\frac{\cal C}{p^2}\left[ \frac{1}{2}\log\left( p^2/\Lambda_2^{\,2} \right) \right]^{\gamma_m-1}\;.
\vspace{0.5mm}
\end{eqnarray}
where the first term is related to the explicit mass in the Lagrangian, $m_R(\mu)$, by some renormalisation factors. The second term gives the lowest dimension vacuum condensate, where ${\cal C}\,=\,-\left< \qq \right>$. 
If we included the expression to all orders then the scales $\Lambda_1$
and $\Lambda_2$ would both be equal to $\Lambda_{QCD}$. However, higher contributions to the leading order forms in Eq.~(\ref{cond:eqn:opefit}) are differently suppressed, and so $\Lambda_1$ and $\Lambda_2$ are in practice different. 
For large masses only the first piece is relevant, whereas for $\overline{m}=0$ only the second term appears. In this latter case, owing to the power suppression of higher orders, $\Lambda_2$ is readily determined and found to be equal to $\Lambda_{QCD}$. Thus in the chiral limit we can easily extract the renormalisation point independent condensate, $\cC\,\equiv\,-\langle \qq \rangle$, from the asymptotics. 

For non-zero current masses,
 one can  attempt to fit both terms of the OPE in Eq.~(\ref{cond:eqn:opefit}) to the tail of the mass function, $M$ of Fig.~\ref{masses}. 
Comparing the full mass with $m_q \ne 0$ with that in the chiral limit, one sees how very small the contribution of the condensate to the tail is. So while a value for the condensate can be extracted, 
this procedure is not at all reliable. 

Strictly in the chiral limit, we may also extract the condensate using:
\begin{equation}\label{cond:eqn:tracecondensate}
\vspace{0.5mm}
	-\left<\qq\right>_\mu = Z_2\left(\mu,\Lambda\right) Z_m\left( \mu,\Lambda \right)N_c\, \tr_D \,\int^\Lambda \frac{d^4k}{\left( 2\pi \right)^4}\,S_F\left(k,\mu\right)\;,
\vspace{0.5mm}
\end{equation}
with $\left<\qq\right>_\mu$ the renormalisation point dependent quark condensate. At one-loop, this is related to the renormalisation point independent one by:
\begin{equation}
\vspace{0.5mm}
	\left<\qq\right>_\mu = \left( \frac{1}{2}\log\frac{\mu^2}{\Lambda^2_{QCD}} \right)^{\gamma_m}\left<\qq\right>\;.
\vspace{0.5mm}
\end{equation}
which we compare with the asymptotic extraction to good agreement.

However, for any non-zero quark masses
, we cannot apply Eq.~(\ref{cond:eqn:tracecondensate}), since it acquires a quadratic divergence, cf. Eq.~(\ref{cond:eqn:opefit}). Indeed, it is the elimination of this divergence that inspired the
definition proposed by Chang {\it et al.}~\cite{Chang:2006bm}, which is unfortunately not equal to the condensate of the physical mass function and  is therefore
ambiguous. Consequently, we need a different definition, one close to the OPE, Eq.~(\ref{cond:eqn:opefit}).

\subsection{Multiple Solutions}
The SDEs, Eq.~(\ref{quark:sde}), can have multiple solutions being as they are non-linear integral equations.
 In the chiral limit, there exist three solutions for $S_F(p)$ and its mass function $M(p^2)$. These correspond to the Wigner mode
, and two non-perturbative solutions of equal magnitude generated by the dynamical breaking of chiral symmetry. These we denote by $M^{\pm,W}(p)$, where:
\begin{equation}\label{cond:eqn:sols} 
	M(p^2) = 	\left\{ \begin{array}{l} 
				M^W(p^2) = 0\;\; \\ [-0.5mm]
                               \\[-0.5mm]
			      M^\pm(p^2) = \pm M^0(p^2) 
			\end{array} \right. \;.
\end{equation}
Analogous solutions to these exist as we move away from the chiral limit, with $M^{-,W}$ restricted to some critical domain $0\le m< m_{cr}$. 

\begin{figure}[t]
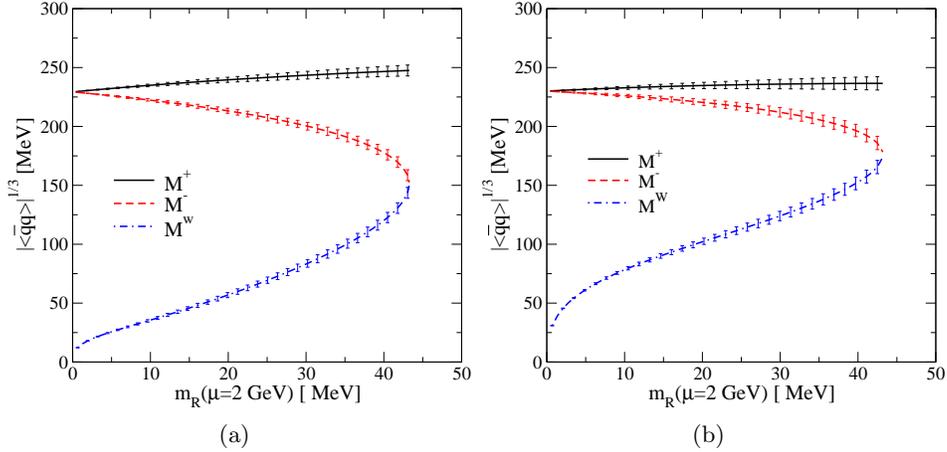

	\begin{center}
	\subfigure[]{\centering\includegraphics*[width=0.49\columnwidth]{polonfig3}\label{cond:fig:cond-lm}}
	\subfigure[]{\centering\includegraphics*[width=0.49\columnwidth]{polonfig4}\label{cond:mt-4-4b}}
	\caption{Condensate extracted through simultaneous fitting of the three solutions to the fermion mass-function in the Maris-Tandy model with $\omega=0.4$ GeV and (a) $N_f=0$, (b) $N_f=4$. The functions of the current quark mass are evolved to 2 GeV in a momentum subtraction scheme.}
	\end{center}
\end{figure}

Instead of one single solution, we now have three solutions to the same model, each with identical running of the current-quark mass (the first term in Eq.~(\ref{cond:eqn:opefit})) in the UV region and differing only by their values of the condensate. Thus, for $m_R(\mu) < m_{cr}$,
it is possible to fit Eq.~(\ref{cond:eqn:opefit}) simultaneously to the three mass functions $M^\pm$, $M^W$. The scales $\Lambda_1$ and $\Lambda_2$ are 
found to be
the same for any current mass within the given model. $\Lambda_2$ is equal to $\Lambda_{QCD}$, while $\Lambda_1$ is roughly twice as big and depends upon the model parameters.
The condensates $\cC^\pm$ and $\cC^W$ are then determined
in an accurate and stable way. 
The results are shown in  Figs.~\ref{cond:fig:cond-lm},~\ref{cond:mt-4-4b} for $N_f=0, 4$. The error bars reflect the accuracy
with which the mass functions  represented by two terms in the OPE expression, Eq.~(\ref{cond:eqn:opefit}), are separable with the anomalous dimensions specified.

We see that within errors the condensate is found to increase with quark mass.
This rise at small masses was anticipated by Novikov {\it et al.}~\cite{Novikov:1981xj} combining a perturbative chiral expansion with QCD sum-rule arguments. That the chiral logs relevant at very small $m_q$ are barely seen is due to the quenching of the gluon and the rainbow approximation of Eq.~(\ref{cond:eqn:quarksde1}). 

We see in Figs.~\ref{cond:fig:cond-lm}, \ref{cond:mt-4-4b} that the $M^-$ and $M^W$ solutions bifurcate below $m_{cr} \simeq 43.4 (44.0)$ MeV with $\omega =0.4$ GeV for $N_f=0(4)$ respectively.
 But what about the value of the condensate for the physical solution $M^+$ beyond the region where $M^-$ and $M^W$ exist, {\it i.e.} $m_R(\mu) > m_{cr}$?
Having accurately determined the scales $\Lambda_1$ and $\Lambda_2$ in the OPE of Eq.~(\ref{cond:eqn:opefit}) in the region where all 3 solutions exist, we
could
 just continue to use the same values in fitting the physical $M^+$ solution alone and find its condensate.
However, this would make
it difficult to produce realistic errors as the quark mass increases.

We see in Figs.~\ref{cond:fig:cond-lm},~\ref{cond:mt-4-4b} that the $M^-$ and $M^W$ solutions bifurcate below $m_{cr}\simeq 43.4(44.0)$ MeV with $\omega =0.4$ GeV for $N_f=0(4)$ respectively. But what about the value of the condensate for the physical solution $M^+$ beyond the region where $M^-$ and $M^W$ exist, \emph{i.e.} $m_R(\mu)>m_{cr}$? Having accurately determined the scales $\Lambda_1$ and $\Lambda_2$ in the OPE of Eq.~(\ref{cond:eqn:opefit}) in the region where all 3 solutions exist, we could just continue to use the same values in fitting the physical $M^+$ solution alone and find its condensate. However, this would make it difficult to produce realistic errors as the quark mass increases.
\begin{figure}[b!]
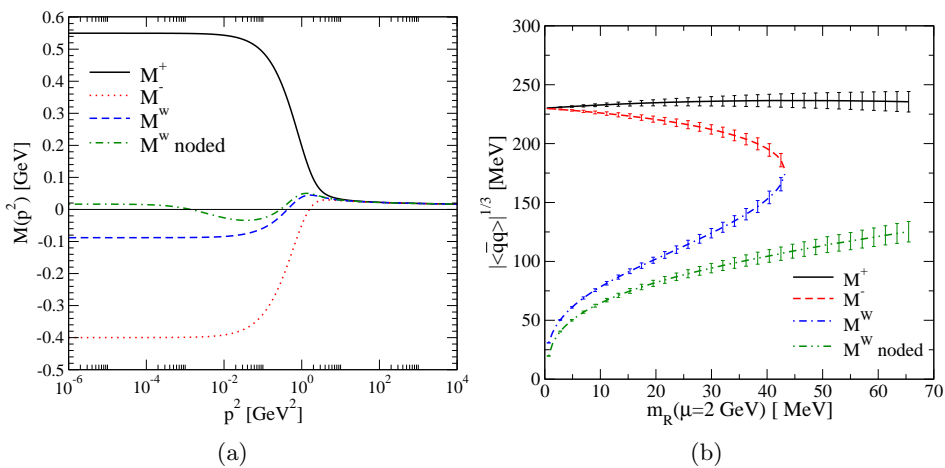

\begin{center}
     \subfigure[]{\includegraphics*[width=0.49\columnwidth]{polonfig5}\label{cond:noded}}
     \subfigure[]{\includegraphics*[width=0.49\columnwidth]{polonfig6}\label{cond:nodecond}}
   \caption{(a) Momentum dependence of the 4 solutions for the fermion
mass-function in the Maris-Tandy model with $m=20$ MeV at $\mu=19$ GeV, and (b) Current quark mass dependence of the condensates, including the noded solution. We use $N_f=4$, $\omega=0.4$ GeV as our parameters.}
\end{center}
\end{figure}

Having relaxed the condition that solutions be positive definite, we in fact find there exist {\it noded} solutions, 
which have also been discovered recently in the context of a simple Yukawa theory
~\cite{Martin:2006qd}. We illustrate this within the Maris-Tandy model, for instance with $N_f=4$ and $\omega=0.4$, in Fig.~\ref{cond:noded}.
\begin{figure}[t]
  \centering
      {\includegraphics*[width=0.62\columnwidth]{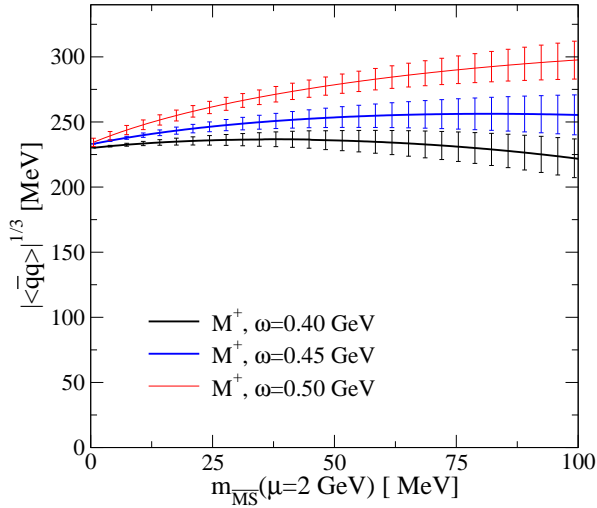}}
	\caption{Condensate for the Maris-Tandy Model with $N_f=4$, $\omega=0.4,0.45,0.5$ GeV as a function of current quark mass defined at 2~GeV in $\overline{MS}$ scheme.}\label{cond:m45MS}
\end{figure} 
We note that this noded solution is not limited to the same domain that restricts $M^-$ and $M^W$.
  Thus there exists a solution with a well-defined UV running of the quark mass exactly as the $M^+$ solution, as far as $m=66.3$ MeV. While at small quark masses we have all four solutions, at larger masses there are still two. Consequently, we can accurately fix
the scales $\Lambda_1$ and $\Lambda_2$ of Eq.~(\ref{cond:eqn:opefit}) at each $m_R(\mu)$, and so determine the condensates as shown in Fig.~\ref{cond:nodecond}.
Indeed, these fits confirm that $\Lambda_1$ and $\Lambda_2$ are independent of $m_R(\mu)$. We can then fit the remaining $M^+$ solutions shown in Fig.~\ref{masses} to give the physical condensate shown in Fig.~\ref{cond:nodecond} for acceptable values of $\omega$ as determined by~\cite{Alkofer:2002bp}. In Fig.~\ref{cond:m45MS} we have in fact scaled the quark mass from $\mu = 2$ GeV in the (quark-gluon) MOM scheme by one loop running to the
$\overline{MS}$ scheme at 2 GeV using the relationship between $\Lambda_{MOM}$ and $\Lambda_{\overline{MS}}$ for 4 flavours deduced by Celmaster and Gonsalves~\cite{Celmaster:1979km}. In this latter scheme the strange quark mass is $\sim 95$ MeV~\cite{PDG}.

\section{Conclusions}
For the range of parameters considered in the Maris-Tandy modelling of strong coupling QCD, we find that the ratio of the condensates at the strange quark mass to the chiral limit to be:
\begin{equation}
	\left<\qq\right>_{m(\overline{MS})\,=95\,{\rm MeV}}/\left<\qq\right>_{m=0}\;=\;(\,1.1\,\pm\,0.2\,)^3,
\end{equation}
in a world with 4 independent flavours. 

What we have shown here is that there is a robust method of determining the value of the $\qq$ condensate beyond the chiral limit based on the Operator Product Expansion. Of course, as the
quark mass increases the contribution of the condensate to the behaviour of the mass function, Fig.~\ref{masses}, becomes relatively less important and so the errors on the extraction of the physical condensate increases considerably.
Nevertheless, the method is reliable up to and beyond the strange quark mass.
Alternative definitions are not.

\section*{Acknowledgements}
RW is grateful to the UK Particle Physics and Astronomy Research Council (PPARC) for the award of a research studentship.
We thank Roman Zwicky and Dominik Nickel for interesting discussions.
This work was supported in part by the EU Contract No. MRTN-CT-2006-035482, 
\lq\lq FLAVIAnet''.

\end{document}